\documentclass[runningheads]{llncs}
\usepackage{graphicx}

\begin{document}
\title{Data Privacy in IoT Equipped Future Smart Homes}
%
%
\author{Athar Khodabakhsh\orcidID{0000-0001-8249-9041}\and
Sule Yildirim Yayilgan\orcidID{0000-0002-1982-6609} 
}
\authorrunning{A. Khodabakhsh and S.Y. Yayilgan}
%
\institute{Department of Information Security and Communication Technology\\
Norwegian University of Science and Technology, NTNU\\
Gjovik, Norway \\
\email{\{athar.khodabakhsh,sule.yildirim\}@ntnu.no}
}
%
\maketitle              
\begin{abstract}

Smart devices are becoming inseparable from daily lives and are improving fast for providing intelligent services and remote monitoring and control. In order to provide personalized and customized services more personal data collection is required. Consequently, intelligent services are becoming intensely personal and they raise concerns regarding data privacy and security. In this paper data privacy requirements in a smart home environment equipped with ``Internet of Things" are described and privacy challenges for data and models are addressed.

\keywords{data privacy \and IoT \and model \and pattern \and privacy \and smart home }
\end{abstract}
\section{Introduction}
\begin{figure}[t]
\begin{center}
\includegraphics[width=3.3in]{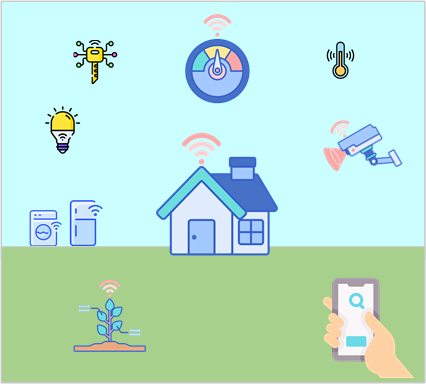}
\caption{Smart home equipped with IoT devices and remote mobile access.} 
\label{fig1}
\end{center}
\end{figure}

Smart home system let the individuals interact with home devices and organize their home intelligently. Heterogeneous electronic smart devices are equipped with sensors, cameras, or actuators and are connected to each other through Internet of Things (IoT) technologies~\cite{privacy:protocol}. These devices are able to collect information from users and home environment to support real-time monitoring, remote control, and safety for smart home. Machine learning algorithms process the collected data and train models to perform personalized and intelligent actions by utilizing techniques such as pattern extraction and speech recognition. Smart service's aim is to provide recommendation to users and bring more intelligence to daily life. Although smart homes bring many advantages in terms of control, management, and cost benefits for users and companies, but also raises concerns about personal data protection, security, and privacy. For a smart service to be reliable \textit{data security and privacy}, \textit{authorization and trust}, \textit{authentication and secure communication}, and \textit{compliance and regulations}~\cite{IoT:farm} should be provided for users. In this paper we discuss data privacy challenges for individuals' activity patterns where data processing gives more insights about their natural behavior~\cite{privacy:anaonymization} therefore, they should have the rights to protect their personal data and be aware of data processing results extracted from their personal information.

Moreover, some IoT devices are produced cheaper and faster by companies to capture the new trend in the market which lead to security risks. Actors with malicious intentions may exploit the vulnerabilities in IoT devices or trained models remotely. They may intrude into smart home's network and analyze internet traffic of smart devices, process user's information, track users' activities, and exploit IoT device vulnerabilities and gradually take over the control of home or lock out the actual residents. 

This paper focuses on data privacy requirements in smart home environment and evolves around a research question: ``what are the requirements to protect personal data and user's profile against privacy violation?" and is organized as follows. Section 2 describes IoT equipped smart home and intelligent services for users. Section 3 explores data privacy challenges and privacy violation scenarios and discusses requirements for smart home privacy policy developments. Section 4 concludes the study and highlights future research.


\section{Smart Home}
Smart home system provides the ability for users to interact remotely with IoT devices via mobile application and finger-print/voice-enabled home automation commands. A smart home consists of several IoT devices depicted in Figure~\ref{fig1} including smart-kitchen devices, smart-meters, smart lighting, smart locks, and wearables~\cite{smartHome:risk}. In Table~\ref{tab1}, their features and functionalities are listed. IoT equipment are interconnected and communicate with each other to provide intelligent services to users. They collect data and user's activities through sensors in order to learn patterns using machine learning techniques.

\begin{table}[t]
\caption{IoT device functionality in smart home.}
\label{tab1}
\begin{tabular}{|l|l|}
\hline
\textbf{~~~~~~~Device} & \textbf{~~~~~~~~~~~~~~~~~~~~~~~~~Functionalities}\\
&\\
\hline
	smart-meters & real-time recording of electricity/water consumption and\\
& interaction with users for consumption patterns\\ \hline
	smart fridge &  flexible user-controlled cooling options and tracking products\\
& that are stored inside \\\hline
	smart-light & remote controlling the light requirements at home with  \\
&customized and scheduled features \\\hline
	surveillance camera & monitoring home environment with motion sensors and \\
&malicious activity detection in area and sending alarms \\\hline
	smart-heating &control and set the environment temperature intelligently \\\hline
	smart-air conditioning& monitoring and customize humidity, smoke, and carbon \\
& monoxide in home environment\\\hline
	smart-key/lock& verified person can enter home or modify system and  \\
&device settings\\\hline
	smart-garden& home growing fresh food and flowers by automated watering, \\
&light, and nutrients\\\hline
	smart-kitchen devices& eco friendly washing machines by reducing water, time, \\
&and energy consumption\\\hline
wearable devices& real-time tracking of the vital signs (via smart watch, etc.)\\\hline
	smart-phone&control smart home system remotely\\ 
\hline
\end{tabular}
\end{table}

\subsection{IoT Intelligent Services}
According to device functionalities described in Table~\ref{tab1}, smart-meter can communicate with smart-heating devices to set the environment temperature, and with smart air-conditioning for setting humidity and smoke level. Smart kitchen devices can provide household services based on resident's presence at home and on holidays based on trained models. Once the pattern is modeled, it can be used for recommendations, generalization to solve problems, estimations and forecasting future requirements. The benefits from smart homes and intelligent customization features can bring automation services such as:
\begin{enumerate}	
\item[$-$] In the smart home system the user can set alarm automatically at a specific time of day when all the residents are away. For instance, this can help to turn off devices that are not used for electricity optimization. 
\item[$-$] The home can be set for warm welcome specially in winter after working hour, including setting the home temperature, boiling water for a drink, etc based on resident's habits and behavior pattern.
\item[$-$] Set up camera to send alarm in case of intrusion utilizing smart and in-built motion sensors and detect frequent visitors through face recognition techniques.
\item [$-$]In smart home it is possible to light up the home before they arrive and turn on/off lights with respect to movement pattern inside home.
\item [$-$]The IoT devices at kitchen such as smart fridge, smart coffee maker, etc. can send notification when an item is finished.
\end{enumerate}

In order to provide these intelligent services, personal data from user's activity are constantly collected and processed for training models and patterns extraction. Personalized models can expose highly sensitive information.

\subsection{Privacy Violation Scenarios}
IoT devices are connected to internet and to each other and having access to one component can lead to direct/indirect access to other smart devices and smart home system. Some of IoT devices are more critical such as smart keys and smart locks. IoT devices collect data about homeowners for better customization and personalization which are stored locally on things or on edge/cloud and unauthorized access to this information can be used for criminal or disruptive activities~\cite{Iot:threats}. Many companies and smart service providers use collected data from IoT devices and train machine learning models to improve advertising and product and service recommendations for users. By using the smart services, activities of the users, their preferences, purchases, health data, transactions, voice commands, and location data are constantly recorded and processed to better understand the data generated by their operations~\cite{attack:mebership}. Personal data collection and records bring two main concerns regarding personal information exposure.

\begin{itemize}
\item \textbf{Attack on data} can expose sensitive information about users:
\begin{itemize}
\item \textit{personal data leakage}: data is less secured when stored on data centers, 
\item \textit{false data injection attack}: attacker might attempt to change/falsify data that is used for real-time decision making, 
\item \textit{misinformation attack}: attacker may release false data reports similar to actual data. 
\end{itemize}
An intruder may gain access to home network, remotely control and exploit sensors and autonomous home devices, track user's activities, and observe smart-cameras in real-time to find out resident's activity pattern and what is going on inside users house~\cite{IoT:farm}. Attackers can use the personal information to unlock the smart keys and turn off alarms during intrusion. Additionally, data processing and analytics are moved to edge for real-time services which gives the attackers an entry point to smart home network through remote-access. 

\item \textbf{Attack on machine learning models} can leak information about the individual data records similar to any other software systems. Privacy attack on machine learning systems such as:
\begin{itemize}
\item \textit{membership interface attacks}: whether a record was in model training dataset,
\item \textit{model inversion attacks}: use a model's output on a hidden input to infer something about this input~\cite{attack:mebership}.
\end{itemize}
For example, training model can uncover high correlations between a user's activities or health features. Once the correlations are known, the information can be used as public facts about the person or members of a population and is a form of privacy breach. Attackers may gain query access to the model and obtain the model’s prediction vector on data records~\cite{attack:mebership}.
\end{itemize}
Therefore, a user's activity pattern can be profiled and may be used to predict aspects concerning natural person's behavior, health, and personal preferences. Personal data is subject to regulatory requirements for protection against violation and should be developed under General Data Protection Regulation (GDPR) standards for novel activity monitoring using machine learning techniques.

\section{Data Privacy}

According to \textit{GDPR, Article.1 Subject-matter and Objectives}: ``regulation to protect fundamental rights and freedoms of natural persons and in particular their right to the protection of personal data" \cite{gdpr} and \textit{GDPR , Article.4 Definitions}:
\begin{itemize}
\item [$Art. 4.2$]Data Processing:``any operation or set of operations which is performed on personal data or on sets of personal data, whether or not by automated means, such as collection, recording, organization, structuring, storage, adaptation or alteration, retrieval, consultation, use, disclosure by transmission, dissemination or otherwise making available, alignment or combination, restriction, erasure or destruction;”\\

\item [$Art. 4.4$] Profiling: ``any form of automated processing of personal data consisting of the use of personal data to evaluate certain personal aspects relating to a natural person, in particular to analyze or predict aspects concerning that natural person’s performance at work, economic situation, health, personal preferences, interests, reliability, behavior, location or movements;"
\end{itemize}
data collection, data processing, and model training which will lead to profiling by applying machine learning techniques are subject to regulations and protection of person's natural behavior pattern.

\subsection{User's Activity Pattern}
For demonstration, we used a public smart home \textbf{UMass Smart* Dataset}, from Smart* project~\cite{umass} for extracting correlation. The goal of Smart* project was to optimize home energy consumption and contains data for 114 single-family apartments for the period 2014-2016. We used Home-A dataset in experiments for pattern extraction.
The dataset includes electricity consumption of 11 devices at smart home such as WashingMachine, FridgeRange, KitchenLights, BedroomLights, MasterLights which are measured every $30~mins$.

\textit{\textbf{Correlation:}} is statistical relationship between two variables and can be positive or negative, where positive correlation means both variables move in same direction and negative value means when one variable increase the other variable decreases~\cite{correlation}. 

\begin{figure}[t]
\begin{center}
\includegraphics[width=3.9in]{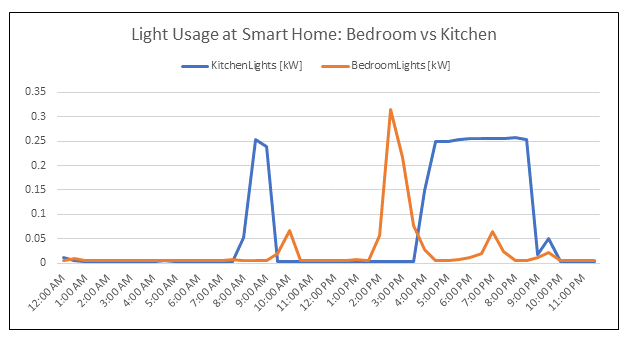}
\caption{Daily light usage at smart home for one day on 1/1/2014.} 
\label{fig2}
\end{center}
\end{figure}

KitchenLights and BedroomLights were selected since have -0.09 correlation. As shown in Figure~\ref{fig2}, the BedroomLights and KitchenLights are not turned on simultaneously and it can be inferred as, the resident turns off the kitchen light when they go to bedroom. Machine learning algorithms can use this knowledge can to extract pattern for electricity consumption management.

\begin{figure}[h!]
\begin{center}
\includegraphics[width=4in]{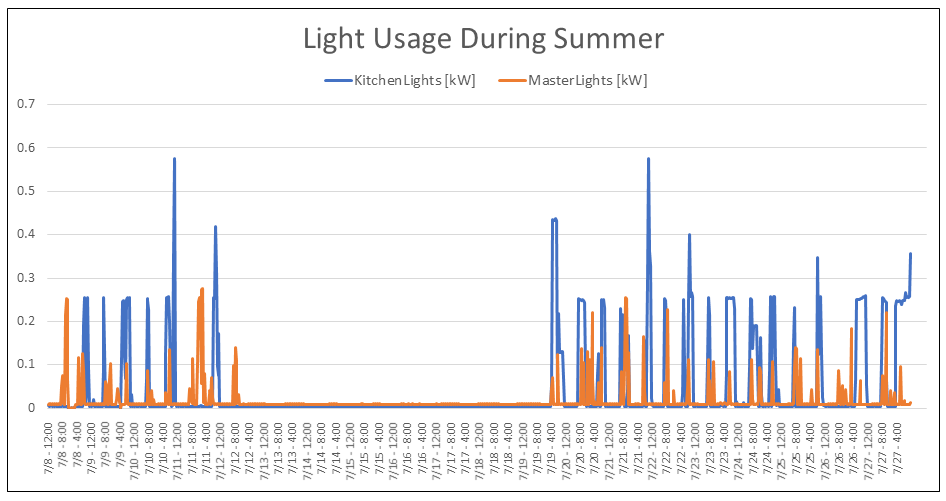}
\caption{Daily light usage at smart home on during 20 days in summer.} 
\label{fig3}
\end{center}
\end{figure}

Another example is correlation between KitchenLights and MasterLights with values of +0.41, which is positive and relatively high correlation. It can be inferred that KitchenLights and MasterLights are turned on simultaneously. In addition, by a quick analysis, absence of the residence can be extracted. As shown in Figure~\ref{fig3}, the lights are not turned on for a period of 5 days by having access to only two features from smart home data. Although this knowledge is very useful for energy optimization, it can be misused if unauthorized actors gain access.

\subsection{Discussion}
Once a natural person's pattern is modeled, it can be used to recommendations, optimization, estimate and forecast requirements by learning relationship among features. Depending on the problem, features are fed to machine learning algorithm for data processing. Security challenges in smart home include \textit{data security and privacy}, \textit{authorization and trust}, \textit{authentication and secure communication}, and \textit{compliance and regulations}~\cite{IoT:farm}. 

%
%

For reliable smart home operation the following points should be considered:
\begin{itemize}
\item Device vendors should be selected based on security quality, data protection, and special features.
\item Devices should have strong passwords and possibly use biometrics for critical information access and settings.
\item The IoT device software and smart home services should be kept updated with latest version.
\item Features that let users for remote access should be turn off if it is not needed.
\item Encryption of all static data must be ensured. 
\item Communications must be encrypted and all the smart devices must be physically secured.
\end{itemize}

\section{Conclusion and Future work}
Artificial intelligence and machine learning are tools to provide smart services however, privacy of natural person and their personal data should be preserved and sensitive training data and their models should be secured. In this paper data privacy challenges in a smart home environment equipped with IoT are addressed. We conclude that mechanisms are required to protect user's and their personal data in smart environments. The lack of security by design in IoT technology can lead to more vulnerabilities and weak security. As future work, we will investigate model protection, device-to device interactions, and data privacy regulations for smart homes.

\end{document}